\documentclass[epjCONF]{svjour}

\usepackage{graphics}
\usepackage[varg]{txfonts} 
\usepackage[latin1]{inputenc}

\session-title{The Space Photometry Revolution (CoRoT Symposium 3, Kepler KASC-7 joint meeting)}

\newcommand{\Msun}{\ensuremath{\,{\rm M}_\odot}}           
\newcommand{\Rsun}{\ensuremath{\,{\rm R}_\odot}}           
\newcommand{\Mjup}{\ensuremath{\,{\rm M}_{\rm Jup}}}       
\newcommand{\Teff}{\ensuremath{T_{\rm eff}}}               
\newcommand{\logg}{\ensuremath{\log g}}                    

\begin{document} 

\title{Multiple star systems observed with CoRoT and Kepler}
\author{John Southworth}
\institute{Astrophysics Group, Keele University, Staffordshire, ST5 5BG, UK}

\abstract{The CoRoT and {\it Kepler} satellites were the first space platforms designed to perform high-precision photometry for a large number of stars. Multiple systems display a wide variety of photometric variability, making them natural benefactors of these missions. I review the work arising from CoRoT and {\it Kepler} observations of multiple systems, with particular emphasis on eclipsing binaries containing giant stars, pulsators, triple eclipses and/or low-mass stars. Many more results remain untapped in the data archives of these missions, and the future holds the promise of K2, TESS and PLATO.}

\maketitle


\section{Introduction} \label{sec:intro}

The CoRoT and {\it Kepler} satellites represent the first generation of astronomical space missions capable of large-scale photometric surveys. The large quantity -- and exquisite quality -- of the data they provided is in the process of revolutionising stellar and planetary astrophysics. In this review I highlight the immense variety of the scientific results from these concurrent missions, as well as the context provided by their precursors and implications for their successors.

CoRoT was led by CNES and ESA, launched on 2006/12/27, and retired in June 2013 after an irretrievable computer failure in November 2012. It performed 24 observing runs, each lasting between 21 and 152 days, with a field of view of $2 \times 1.3^\circ \times 1.3^\circ$, obtaining light curves of 163\,000 stars \cite{Moutou13icar}. {\it Kepler} was a NASA mission, launched on 2009/03/07 and suffering a critical pointing failure on 2013/05/11. It observed the same 105\,deg$^2$ sky area for its full mission duration, obtaining high-precision light curves of approximately 191\,000 stars. The remainder of the spacecraft is fully functional and has now been reincarnated as the K2 survey instrument, which is sequentially observing ten fields on the ecliptic for approximately 75 days each with a reduced photometric precision \cite{Howell14pasp}.

The high-precision photometric capabilities of CoRoT and {\it Kepler} mean their primary contribution to the study of multiple stars is obtaining light curves of eclipses, plus proximity-induced phenomena such as the reflection, ellipsoidal and Doppler beaming effects. Intrinsic stellar variability such as pulsations are also detected in abundance, including for many components of multiple systems. This review is a tree-tops perspective on the forest of results harvested from the CoRoT and {\it Kepler} data.

\section{Eclipses in binary stars} \label{sec:ecl}

\begin{figure}
\resizebox{\columnwidth}{!}{\includegraphics{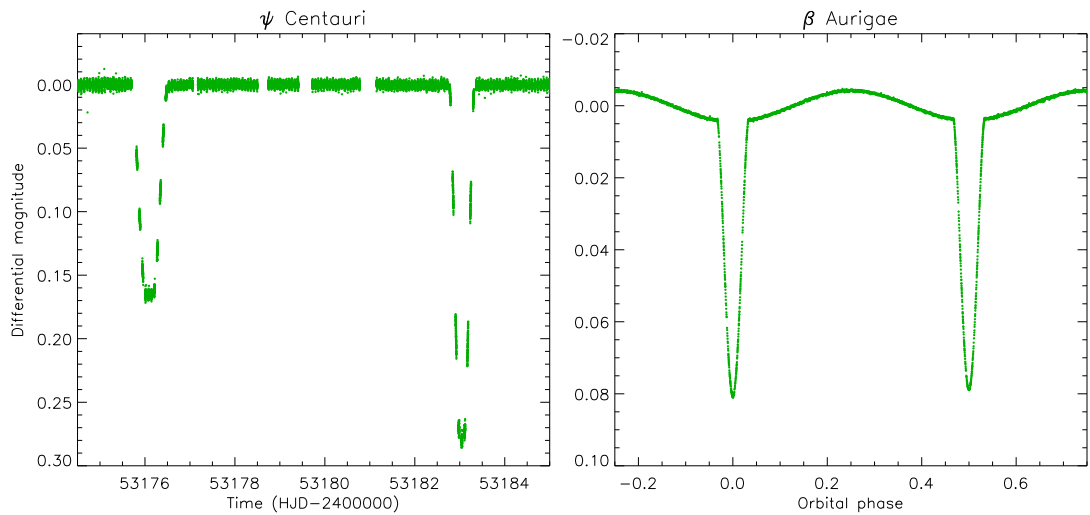}}
\caption{WIRE satellite light curves of the eclipsing systems $\psi$\,Centauri (left) and $\beta$\,Aurigae (right).}
\label{fig:1}
\end{figure}

The importance of eclipsing binaries (EBs) is that it is possible to measure the masses and radii of their components directly from light and radial velocity curves, and to high precision (e.g.\ \cite{me05mn}). Measurement of the \Teff\ values of the stars enables luminosity to be calculated directly, making them excellent distance indicators (e.g.\ \cite{me05aa,Graczyk14apj}). These properties are useful in critiquing predictions from thereotical stellar models \cite{Pols97mn,Pavlovski09mn} and model atmospheres \cite{Howarth11mn}, investigating star and binary formation scenarios \cite{Albrecht09nat} and galactic chemical evolution \cite{Ribas00mn}, establishing empirical relations for use in stellar physics \cite{Torres10aarv}, and determining the properties of transiting extrasolar planets (e.g.\ \cite{me09mn,Gillon14aa}).

The study of EBs is a mature field: the term ``binary star'' was christened by William Herschel in 1802 \cite{Herschel1802}. Eclipses were proposed by John Goodricke in 1783, as a possible explanation for $\beta$\,Persei (Algol), whose striking periodic variability may have been noted as early as 1672 by Geminiano Montanari \cite{French12javso}. Vogel proved the binary nature of $\beta$\,Persei through spectroscopy \cite{Vogel1890pasp}, and Stebbins attained the first direct measurement of the masses and radii of stars in an EB, $\beta$\,Aurigae \cite{Stebbins11apj}.

Space-based observations of EBs also date back surprisingly far, the enigmatic object $\beta$\,Lyr being a target for the {\it Orbiting Astronomical Observatory} (OAO-2) in 1970, and for the Voyager\,1 and Voyager\,2 ultraviolet spectrometers in the 1980s \cite{Kondo94apj}. The International Ultraviolet Explorer (IUE) obtained light curves of over 300 EBs, although most are very sparse \cite{Polubek03}. A complete light curve of VV\,Orionis was obtained by OAO-2 in 1969 and 1970 \cite{Eaton75apj}.

The first of the modern era of high-precision light curves of EBs was gathered by the star tracker on the WIRE satellite in 2004 for $\psi$\,Cen \cite{Bruntt06aa}, a serendipitous result as this nearby star was not previously know to be binary (let alone eclipsing) and was automatically chosen as a comparison star for another object by onboard software. The WIRE star tracker was also used to obtain a remarkable light curve of $\beta$\,Aurigae in 2006 \cite{me07aa}. The WIRE photometry for $\psi$\,Centauri and $\beta$\,Aurigae are shown in Fig.\,1. Observations of a small number of eclipsing systems have also been made using the MOST satellite (e.g.\ \cite{Rucinski07mn,Pribulla10an,Windemuth13apj}).

The overall picture before the launch of CoRoT and {\it Kepler} is therefore one where only a handful of multiple systems had been the subject of detailed study from space, using telescopes optimised for observing a single target at a time. I now turn to the results achieved with CoRoT and {\it Kepler}, satellites capable of observing thousands of stars simultaneously.


\section{The CoRoT and {\it Kepler} era} \label{sec:ck}

These two satellites have allowed significant advances to be made for a wide variety of stellar types and physical phenomena. Below I highlight some of the most influential and intriguing results conerning multiple stars, made possible by the unprecedented quality and quantity of the data harvested by CoRoT and {\it Kepler}.

\subsection{Giant stars in eclipsing binaries} \label{sec:ck:giant}

Arguably the most valuable EBs are well-detached: those whose components' evolution has been negligibly affected by the presence of their companion are more representative of single stars and best suited to comparision with single-star evolutionary theory. This presents a problem in the study of evolved stars in EBs, because their swollen radii mean orbital periods must be long for them to remain in a detached state. This in turn makes photometric follow-up extremely expensive in terms of telescope time and human effort. Prior to CoRoT and {\it Kepler} the only well-measured detached EB with a component with $\logg < 3$ was TZ\,For, a 75-d period system studied using a semi-autonomous photometer \cite{Andersen91aa}.

{\it Kepler} has been used to discover many EBs containing giants. The first of these, KIC 8410637, showed a total eclipse of depth 0.1\,mag and duration 2.2\,d in the Q0 data \cite{Hekker10apj}. As further quarters of data accumulated, a secondary eclipse of depth 0.03\,mag and duration 8.1\,d materialised. Further eclipses from {\it Kepler} established its orbital period as 408\,d. In combination with RV follow-up from terrestrial telescopes, these data enabled measurement of the physical properties of the system \cite{Frandsen13aa}. The more massive star is an 11\Rsun\ giant with a \Teff\ of 4800\,K, showing stochastic oscillations, whereas its companion is a 6500\,K F-dwarf. A detailed study of this oscillating giant of precisely known mass and radius will enable a check on the predictive power of asteroseismology. A further 12 promising candidates have been found in the {\it Kepler} dataset \cite{Gaulme13apj}; the stochastic oscillations in those with shorter orbital periods appear to be damped \cite{Gaulme14apj}.

\begin{figure}
\resizebox{\columnwidth}{!}{\includegraphics{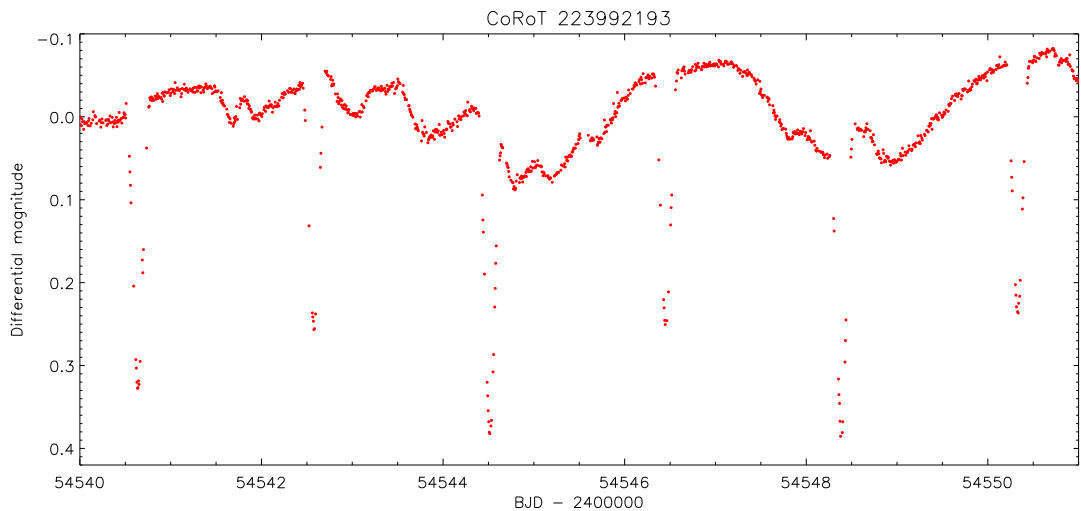}}
\caption{Excerpt from the CoRoT 512s-cadence light curve of the pre-main-sequence eclipsing binary CoRoT 223992193.}
\label{fig:223992193}
\end{figure}

\subsection{Multiply-eclipsing systems} \label{sec:ck:multi}

One of the most critical benefits of near-continuous high-precision photometry is the ability to detect -- and study in detail -- unusual and particularly interesting objects. KOI-126 turned out to be a triple system composed of a short-period EB containing two 0.2\Msun\ stars, itself eclipsing and being eclipsed by a slightly evolved G-star (mass 1.3\Msun, radius 2.0\Rsun, \Teff\ 5875\,K) \cite{Carter11sci}. Although the two M-dwarfs were spectroscopically undetectable, the system could be characterised to high precision (masses to 1\%, radii to 0.5\%) by modelling the continually-changing duration, depth and complex shapes of the mutual events. This approach required a `photodynamical' model which twinned the calculation of eclipse light loss with an $N$-body code to account for the dynamical effects.

HD\,181068 (KIC 5952403), a bright star observed with {\it Kepler}, exhibits short and shallow eclipses on a 0.9\,d period attributable to an EB containing two K-dwarfs, which in turn transit and occult a G-giant (mass 3.0\Msun, radius 12.5\Rsun) on a 42\,d period \cite{Derekas11sci}. The very unusual eclipse morphology for this object, where the short/shallow eclipses disappear during only alternate long/deep eclipses, is caused by the very similar \Teff s of the three stars (5100\,K for the giant, and 5100\,K and 4675\,K for the dwarfs) \cite{Borkovits13mn}. Another stunning example of tertiary eclipses is displayed by KIC 2856960 \cite{Armstrong12aa}.

\subsection{Low-mass and pre-main-sequence systems} \label{sec:ck:vlm} \label{sec:ck:pms}

The study of low-mass stars in EBs is vital because theoretical descriptions of these objects mismatch observed quantities: they persistently show radii larger than expected for their mass (see \cite{Torres13an} for a recent review). This line of research, however, is hindered by the sparsity of potential targets (e.g.\ \cite{Irwin14xxx}). KIC 6131659 is an example with a relatively long orbital period \cite{Bass12apj}; it shows a good agreement with theoretical models, implying that tidally-induced magnetic effects are the culprit of the inflated radii of low-mass stars in EBs. A further 200 low-mass EBs have been identified using {\it Kepler} data \cite{Coughlin11aj}, but results are yet to appear for these systems.

Pre-main-sequence stars are short-lived objects for which theoretical models are very poorly constrained. CoRoT was enlisted to help this issue through observations of the young open cluster NGC 2264, which is known to contain many young stars. Several tens of EBs were identified, and CoRoT 223992193 has been studied in detail \cite{Gillen14aa}. Its light curve shows well-defined eclipses plus large and slower stochastic variations attributed to the occultation of material near the inner edge of a circumbinary disc (Fig.\,\ref{fig:223992193}). The masses and radii of the two stars have been measured precisely and provide important constraints in a previously empty tract of the mass--radius diagram.

\subsection{$\delta$ Scuti stars in eclipsing binaries} \label{sec:ck:dsct}

\begin{figure}
\resizebox{\columnwidth}{!}{\includegraphics{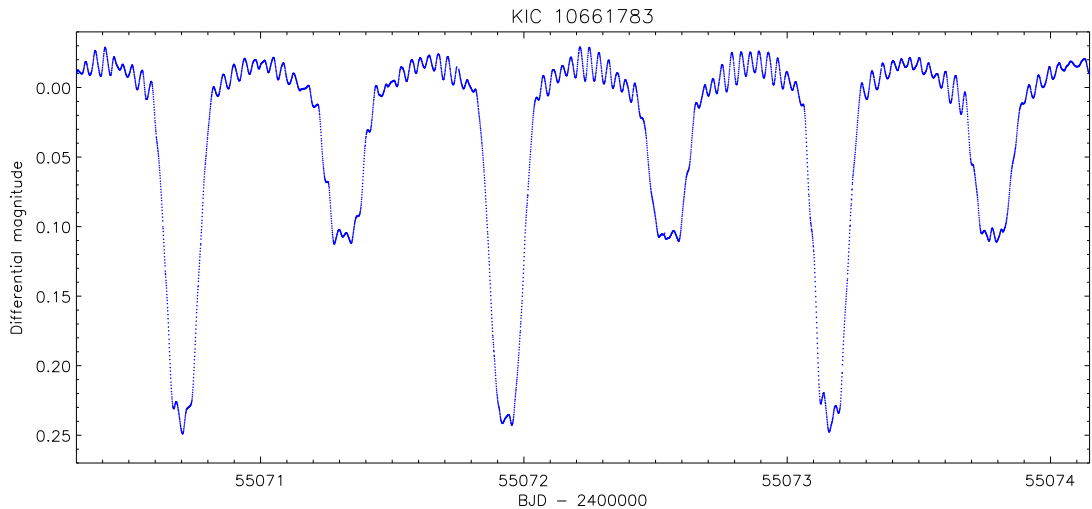}}
\caption{Excerpt from the {\it Kepler} short-cadence light curve of the $\delta$\,Scuti eclipsing system KIC 10661783.}
\label{fig:10661783}
\end{figure}

$\delta$\,Scuti stars are a relatively familiar class of pulsating A-type stars, showing periodic radial and non-radial pulsations at frequencies 3--70\,d$^{-1}$ \cite{Grigahcene10apj}. A significant number have been observed in EBs using CoRoT and {\it Kepler}. An early result was for KIC 10661783, a totally-eclipsing Algol system whose primary component shows over 50 pulsational frequencies in the range 18--31\,d$^{-1}$ \cite{me11mn} (see Fig.\,\ref{fig:10661783}). Follow-up spectroscopy \cite{Lehmann13aa} facilitated the measurement of masses and radii precise to 1\%, which are inconsistent with the mass ratio required to fit the light curve if the secondary star fills its Roche lobe. It seems that the star is a detached post-Algol system. Definitive conclusions await the construction of more sophisticated light curve synthesis codes, which are sufficiently realistic to cope with data of the quality provided by {\it kepler}.

CoRoT 105906206 is a detached EB containing a $\delta$\,Scuti star, for which precise masses and radii were measured \cite{Dasilva14aa}. This process required the inclusion of Doppler beaming \cite{Loeb03apj} in the light curve model. Another lovely example of such systems is KIC 3858884, with a 26.0-day eccentric orbit ($e = 0.47$) and a secondary star showing oscillations in multiple high-order $g$-modes \cite{Maceroni14aa}.

The promise of working on $\delta$\,Scuti stars in EBs is the possibility to perform mode identification, enabled by precise knowledge of the mass and radius of the pulsating star \cite{Creevey11apj}. This is notoriously tricky to achieve, but would specify precise constraints on stellar theory. Another possibility enabled by eclipses is obtaining spatial information for the pulsations on the surface of the star \cite{Biro11mn}.

\subsection{$\gamma$ Doradus stars in eclipsing binaries} \label{sec:ck:gdor}

\begin{figure}
\resizebox{\columnwidth}{!}{\includegraphics{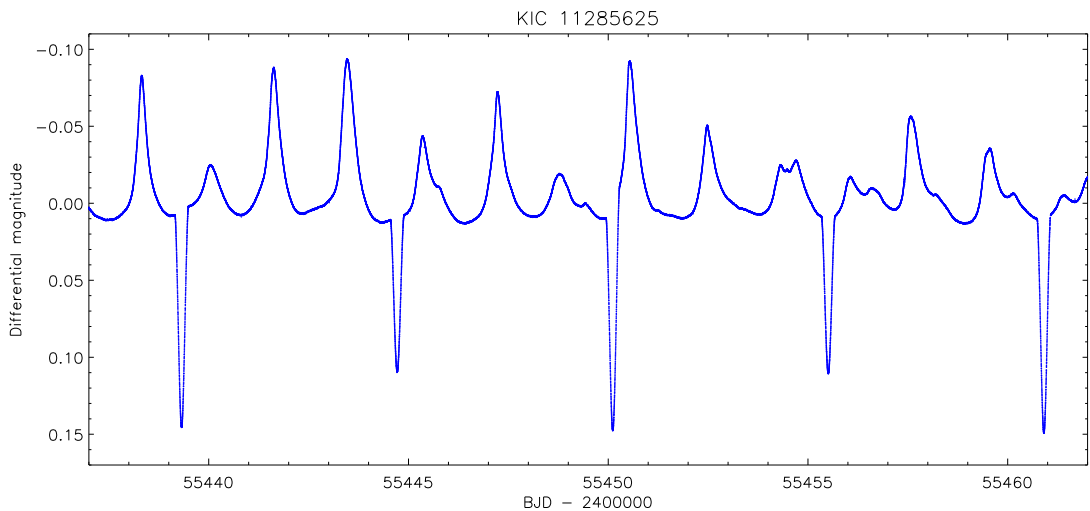}}
\caption{Excerpt from the {\it Kepler} long-cadence light curve of the $\gamma$\,Doradus eclipsing system KIC 11285625.}
\label{fig:11285625}
\end{figure}

CoRoT 102918586 is a detached EB whose primary component is an example of the $\gamma$\,Dor variability class -- late-A to early-F dwarfs which show high-order non-radial $g$-mode pulsations with typical frequencies 0.3-3\,d$^{-1}$ \cite{Handler99mn}. The CoRoT light curve and ground-based \'echelle spectroscopy yielded mass and radius measurements to 1\% precision and accuracy \cite{Maceroni13aa}. The pulsations are consistent with high-order $\ell = 1$ $g$-modes for the measured mass and radius of the primary star.

KIC 11285625 is a spectacular example of a pulsating star in an EB (Fig.\,\ref{fig:11285625}), showing pulsations of up to 0.1\,mag amplitude overlaid with 0.15\,mag deep eclipses on a 10.8\,d period \cite{Debosscher13aa}. By twinning a simple light curve code \cite{me04mn} with pulsational modelling, \cite{Debosscher13aa} measured the physical properties of the stars, and found amplitude modulation and frequency splittings at the orbital and rotational periods.

Hybrid $\delta$\,Sct\,/\,$\gamma$\,Dor stars exist in the region of overlap between the pulsation classes in the {Hertz}-{sprung}-Russell diagram \cite{Grigahcene10apj}. Such objects are also apparent in EBs, for example CoRoT 100866999 \cite{Chapellier13aa}. KIC 4544587 also shows both $p$-mode and $g$-mode pulsations superimposed on deep eclipses in an eccentric orbit \cite{Hambleton13mn}; in this case it is likely that the primary star is a $\delta$\,Sct and the secondary star is a $\gamma$\,Dor. Such a situation highlights the difficulty in assigning oscillation frequencies to individual components of an unresolved binary.

\subsection{Heartbeat stars and tidally-induced pulsations} \label{sec:ck:heartbeat}

\begin{figure}
\resizebox{\columnwidth}{!}{\includegraphics{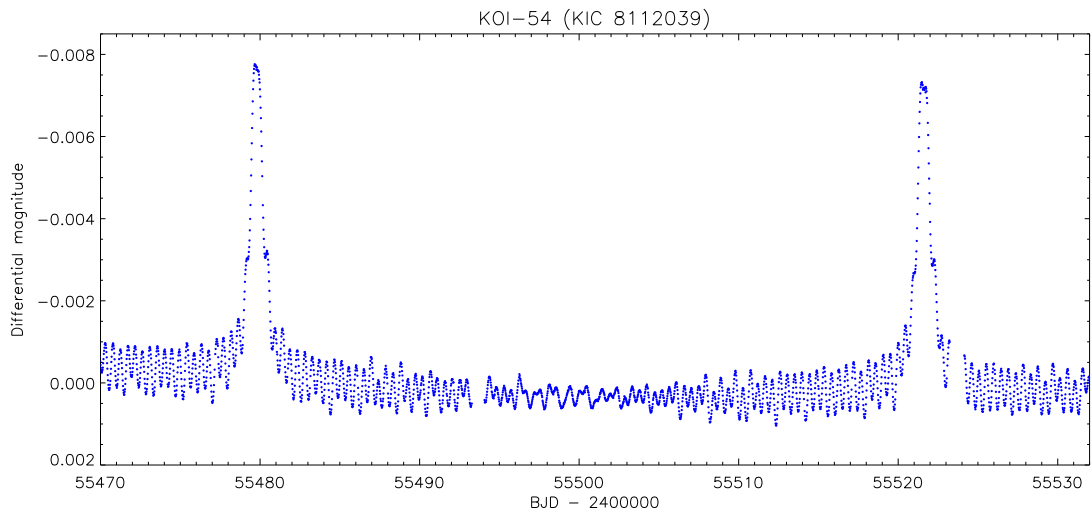}}
\caption{Excerpt from the {\it Kepler} long-cadence light curve of the heartbeat star KOI-54.}
\label{fig:koi54}
\end{figure}

KOI-54 shows an extremely unusual light curve (Fig.\,\ref{fig:koi54}) containing 0.6\% brightenings every 42\,d and obvious oscillations at the 90\hspace{1pt}th and 91\hspace{1pt}st multiple of the corresponding frequency \cite{Welsh11apjs}. With the aid of spectroscopic radial velocity measurements the brightening was explained as resulting from the ellipsoidal and irradiation effects during periastron passage of two A-stars in a highly eccentric ($e = 0.83$) binary system. Over 30 oscillation periods were detected at integer multiples of the orbital frequency ranging from 23 to 91 $f_{\rm orb}$, a clear indication of tidal excitation. Over 100 of these objects are now known, and some are also eclipsing (\cite{Thompson12apj}, K.\ M.\ Hambleton, priv.\,comm.).

Tidally-induced pulsations have been seen before in several EBs. The first clear detection was from the CoRoT light curve of HD\,174884, which has an orbital period of 3.6\,d and oscillations at 8 and 13 $f_{\rm orb}$ \cite{Maceroni09aa}. HD\,181068 (see Section \ref{sec:ck:multi}) also shows oscillations which are probably tidally induced \cite{Derekas11sci}.

\subsection{Stochastic oscillations in V380\,Cygni} \label{sec:ck:v380}

Stochastic (solar-like) oscillations are a fundamental component of modern stellar astrophysics, due in large part to the capabilities of CoRoT and {\it Kepler}. They were an integral part of the strategy for characterising the transiting planets discovered using {\it Kepler} (e.g.\ \cite{Christensen10apj,Marcy14apjs}) and will repeat this role for the forthcoming PLATO mission \cite{Rauer13xxx}. The goal of validating asteroseismic results, using empirically-measured properties of stars in EBs, has not yet been possible due to a lack of available targets. This is because tidal effects in many EBs cause them to rotate faster, thus suppressing stochastic oscillations.

Stochastic oscillations have, however, been detected in the more massive and evolved (11.4\Msun, 15.7\Rsun) component of the eccentric EB V380\,Cyg \cite{Tkachenko12mn}. A detailed analysis of this system \cite{Tkachenko14mn} found that evolutionary models for massive stars predict insufficient mixing near the convective core.

\subsection{Circumbinary planets} \label{sec:ck:cbp}

The possibility of planets orbiting binary stars has a long history, especially in science fiction. Whilst there have been several claims based on eclipse timing variations, the first unarguable case was that of Kepler-16 \cite{Doyle11sci}. This system consists of a 41-d EB with 0.7\Msun\ and 0.2\Msun\ components orbited by a 0.33\Mjup\ transiting planet. The timing and duration of the transits vary greatly due to the orbital motion of the stars in the binary. At the time of writing ten transiting circumbinary planets are known, in eight systems, the most recent being KIC 9632895 \cite{Welsh14xxx}. A huge advantage of these objects is that dynamical interactions affect the transits, offering a way to measure the masses of the stars and planets to sometimes high precision without recourse to extensive spectroscopic follow-up observations. In the case of Kepler-16, the stellar masses and radii are known to 0.5\%, and the planetary mass and radius to 5\% and 0.3\%.

\subsection{Interacting binaries} \label{sec:ck:ib}

\begin{figure}
\resizebox{\columnwidth}{!}{\includegraphics{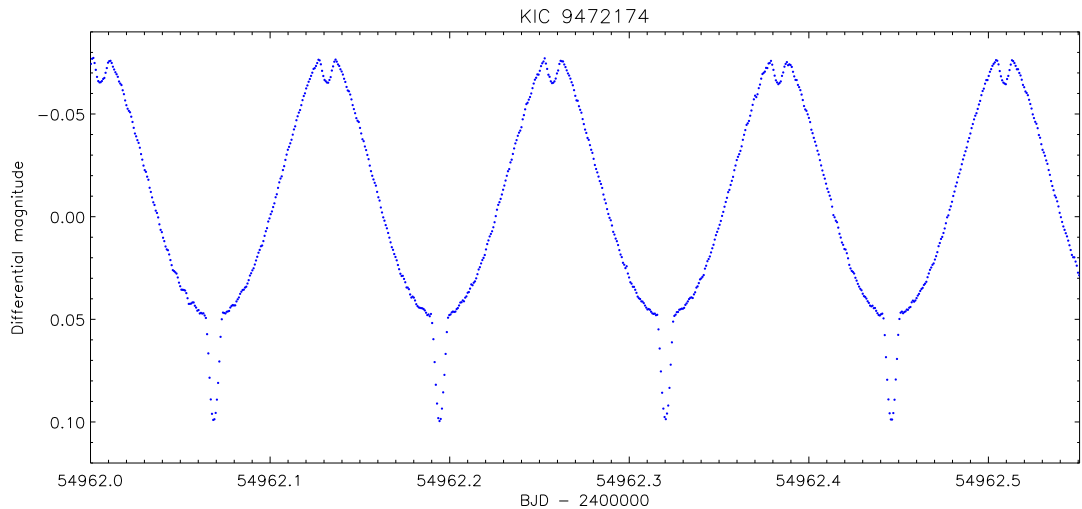}}
\caption{Excerpt from the {\it Kepler} short-cadence light curve of the sdB plus M-dwarf eclipsing binary KIC 9472174.}
\label{fig:9472174}
\end{figure}

Whilst this review has concentrated on eclipses and pulsations, binary-star evolution gives rise to myriad phenomena which cause photometric variability. Some of the most curious arise from cataclysmic variables, objects containing a white dwarf accreting material from a Roche-lobe-filling low-mass star. These objects variously show outbursts, superoutbursts, eclipses, superhumps, novae, pulsations in the white dwarf, and/or long periods of enhanced or diminished brightness. The {\it Kepler} light curve of V344\,Lyr showed five outbursts and one superoutburst over three months \cite{Still10apj}. The almost-continuous monitoring and high data quality enabled detection of sumperhumps both during outburst and quiescense. Another system, V477\,Lyr, showed outbursts, superoutbursts and eclipses \cite{Ramsay12mn}.

KPD\,1946+4340 is an EB containing a subdwarf B (sdB) star and a white dwarf \cite{Bloemen11mn}; obtaining a good fit to the {\it Kepler} light curve required the inclusion Doppler beaming. KIC 9472174 is an eclipsing sdB plus M-dwarf system showing a strong reflection effect, shallow primary and secondary eclipses (Fig.\,\ref{fig:9472174}), and a rich $p$- and $g$-mode spectrum arising from the sdB star \cite{Ostensen10mn}.


\section{Outlook and future missions} \label{sec:future}

Although the CoRoT and main {\it Kepler} missions have both concluded due to hardware issues, their archives hold a huge amount of untapped data. The current version of the {\it Kepler} eclipsing binary catalogue \cite{Prsa11aj} contains over 2000 objects of which only a small fraction have been subjected to detailed analysis. The primary bottleneck to this work is a lack of manpower. Both missions therefore continue to yield new results which further our understanding of multiple and single stars.

Space-based photometry continues to be available from several missions of more limited scope. {\it Kepler} itself continues to observe at reduced photometric precision in its reincarnation as the K2 mission, and has already observed a significant number of EBs \cite{Conroy14xxx}. Most of the BRITE constellation of small satellites \cite{Weiss14pasp} have now been launched and are returning data. BRITE will obtain multi-colour light curves of some of the brightest objects in the sky, although the number of targets is low due to limitations on the rate at which data can be transferred to the ground.

The next landmark mission is TESS \cite{Ricker14spie}, which will photometrically observe 26 fields covering most of the sky. TESS will solve some of the acknowledged problems of {\it Kepler} (a limited field of view containing relatively faint stars) but at the expense of coarser spatial resolution and much shorter light curve durations (27 days near the ecliptic ranging to one year around the celestial poles). It is slated for launch in 2017 and will observe for two years with the possibility of a mission extension.

Further ahead, the ESA PLATO mission \cite{Rauer13xxx} is planned for launch in 2024 as a precision photometry survey instrument. PLATO will solve the {\it \'etendue conundrum} by consisting of 32 small telescopes on a common platform, enabling it to achieve a high photometric precision on many bright stars over a large field of view. An additional huge advantage is onboard data reduction, which will enable higher time resolution compared to CoRoT and {\it Kepler} (25\,s versus mostly 512\,s or 1765\,s sampling) and allow data to be obtained for all point sources in the field of view. The author will lead the eclipsing binary working group for PLATO; please contact me if you wish to be involved.

In conclusion, CoRoT and {\it Kepler} have caused a fundamental change in the study of multiple stars by providing light curves of unprecedented quality and quantity. An impressive number and variety of results have been gathered, and the immense databases from these missions will power stellar and planetary physics for many more years. Work remains to be done on our understanding of high-mass and low-mass stars, on asteroseismology via pulsating stars in EBs, on the effective temperature scale, and on progressively finer tests of the predictions of theoretical models.


\end{document}